\documentclass[twoside,reqno]{article}
\usepackage{epsfig,cite}
\usepackage{amssymb,amsmath}
\newpage
\def \bea {\begin{eqnarray}}      
\def \mea {\nonumber\\}    
\def \eea {\end{eqnarray}}      

\begin{document}    
\begin{titlepage}     
\title{What is semiquantum mechanics?}

\author{A.J. Bracken\footnote{Talk presented at the IVth International Symposium on Quantum Theory and Symmetries,
Varna, August, 2005. {\em Email:} ajb@maths.uq.edu.au} \\Department of Mathematics\\    
University of Queensland\\Brisbane 4072\\Queensland\\Australia}     
     
\date{}     
\maketitle     
     
\begin{abstract}







Semiclassical approximations to quantum dynamics are almost as old as quantum mechanics itself.  In the approach pioneered by Wigner,
the evolution of his quasiprobability density function on phase space is expressed as an asymptotic series 
in increasing powers of Planck's constant, with the classical Liouvillean evolution as leading term. Successive
semiclassical approximations to quantum dynamics are defined by successive terms in the series.
We consider a complementary approach,
which explores the quantum-clssical interface from the other direction.  Classical dynamics is formulated in Hilbert space, 
with the Groenewold quasidensity operator as the image of the Liouville density on phase space.  The evolution of the 
Groenewold operator is then expressed as an asymptotic series in increasing powers of Planck's constant. 
Successive semiquantum approximations to classical dynamics are defined by successive terms in this series, with the familiar
quantum evolution as leading term.    
\end{abstract}
\end{titlepage}


\section{Introduction}
Wigner introduced his famous quasiprobability density function
on phase space in order to consider semiclassical approximations to the quantum evolution of 
the density matrix \cite{Wigner32}.  For   a system with one degree of freedom and classical Hamiltonian
\bea
H=p^2/2m +V(q)\,,
\label{hamiltonian1}
\eea
he found for the evolution of the Wigner function $W(q,\,p,\,t)$,
\bea
W(q,\,p,\,t)_t
=V'(q)\,W(q,\,p,\,t)_p
-\frac{p}{m}\,W(q,\,p,\,t)_q
\mea\mea
-\frac{\hbar^2}{24}\left\{
V'''(q)W(q,\,p,\,t)_{ppp}
-\frac{3p}{m}\,V''(q)W(q,\,p,\,t)_{qpp}\right.
\mea\mea
\left. +\frac{3}{m}\,V'(q)W(q,\,p,\,t)_{qqp}\right\}+{\rm O}(\hbar^4)\,. 
\label{wigner_evolution1}
\eea
Here we have introduced the notation $V'(q)$ for $dV(q)/dq$ and $W_{p}$ for 
$\partial W/\partial p$, {\em etc.}  The first two terms on the RHS in
(\ref{wigner_evolution1}) define the classical Liouvillean evolution, while the 
terms of order $\hbar^2$ define the first semiclassical approximation to the full quantum evolution 
of the Wigner function, and so on. 

Following the subsequent works of Groenewold \cite{Groenewold46} and Moyal \cite{Moyal49},
we now recognize the RHS of (\ref{wigner_evolution1}) as the expansion in ascending powers
of $\hbar$ of the star, or Moyal, bracket of the Hamiltonian $H$ and the Wigner function $W$,
\bea
\{H,\,W\}_{\star}
=
\frac{2}{\hbar}\,H\,\sin\left(\hbar J/2\right)\,W\,,\qquad
J=
\frac{\partial^L}{\partial q}
\frac{\partial^R}{\partial p}
-
\frac{\partial^L}{\partial p}
\frac{\partial^R}{\partial q}\,,
\mea\mea
W_t=\{H,\,W\}_{\star}=H\,J\,W-\frac{\hbar^2}{24}\,H\,J^3\,W+{\rm O}(\hbar^4)\,.\qquad
\label{star_bracket1}
\eea
Here the superscripts $R$ and $L$ in the Janus operator $J$ indicate the directions in which the
differential operators act.  The leading term $H\,J\,W$ in the last equation 
represents the Poisson bracket of $H$ and $W$, and corresponds to the 
first two terms on the RHS in (\ref{wigner_evolution1}).  

All this is very well known.  It is a central ingredient of the so-called phase space
formulation of quantum mechanics \cite{Dubin00,Zachos02}, where operators on Hilbert space are 
mapped into functions on phase space, and in particular the density operator is mapped into
the Wigner function. 

Less well known is that, in a completely analogous way, classical mechanics 
can be reformulated in Hilbert space \cite{Groenewold46,Vercin00,Bracken03}, with the
classical Liouville density mapped into a quasidensity operator \cite{Muga92,Muga93,Muga94} that we
have called elsewhere \cite{Wood04} the Groenewold operator.  The evolution of
this operator in time is defined by what we have called \cite{Bracken03}
the odot bracket, and the expansion of this bracket in ascending powers of
$\hbar$ defines a series of semiquantum approximations to classical dynamics, starting
with the quantum commutator.  

In this way, we explore the classical-quantum interface in a new way, approaching classical 
mechanics from quantum mechanics, which is now regarded as a first approximation.  So we stand on its head the
traditional approach, which approaches quantum mechanics from classical mechanics, regarded as a first approximation.

\section{The Weyl-Wigner transform}

In order to see how this works, we recall firstly \cite{Dubin00,Zachos02,Cassinelli03} that the 
phase space formulation of quantum mechanics is defined
by the Weyl-Wigner transform ${\cal W}$, which maps operators ${\hat A}$ on Hilbert space
into functions $A$ on phase space,
\bea
A={\cal W}({\hat A})\,,\qquad A=A(q,\,p)\,,
\label{WW_transform1}
\eea
and in particular defines the Wigner function
$W={\cal W}({\hat \rho}/2\pi\hbar)$,
where ${\hat \rho}$ is the density matrix defining the state of the quantum system.  
Then
\bea
\langle {\hat A}\rangle(t)={\rm Tr}({\hat \rho}(t) {\hat A})=\int A(q,\,p)\,W(q,\,p,\,t)\,dq\,dp=\langle A\rangle (t)\,,
\mea\mea
\int W(q,\,p,\,t)\,dq\,dp=1\,,\qquad\qquad\qquad
\label{averages1}
\eea
but $W$ is not in general everywhere nonnegative; it is a quasiprobability density function.

In more detail, $A$ is defined by first regarding ${\hat A}$ as an 
integral operator with kernel $A_K(x,\,y)=\langle x|{\hat A}|y\rangle$
in the coordinate representation,  
and then setting 
\bea
A(q,\,p)={\cal W}({\hat A})(q,\,p)=\int     
A_K(q-x/2,\,q+x/2)\,e^{ipx/\hbar}\,dx\,.     
\label{WW_transform2}
\eea
The transform of the operator product on Hilbert space then 
defines the noncommutative star product
on phase space,
\bea
{\cal W}({\hat A}{\hat B})=A\star B\,,
\label{star_product}
\eea
leading to the star bracket as the image of $(1/i\hbar \times)$ the commutator,
\bea
{\cal W}([{\hat A},\,{\hat B}]/i\hbar)=(A\star B-B\star A)/i\hbar=\{A,\,B\}_{\star}\,.
\label{star_bracket}
\eea 
It can now be seen that the quantum evolution of the density matrix
\bea
{\hat \rho}_t= \frac{1}{i\hbar}[{\hat H},\,{\hat \rho}]\,,
\label{quantum_evolution}
\eea 
is mapped by the Weyl-Wigner transform ${\cal W}$ into the evolution
equation for the Wigner function
\bea
W_t=\{H,\,W\}_{\star}
\label{wigner_evolution3}
\eea
as in (\ref{star_bracket1}), so leading to the sequence of semiclassical approximations as described in the Introduction. 
In (\ref{quantum_evolution}), ${\hat H}$ is the quantum Hamiltonian operator, so that $H={\cal W}({\hat H})$.   

In order to define semiquantum approximations to classical dynamics, we begin by considering the
inverse Weyl-Wigner transform ${\cal W}^{-1}$, which maps functions $A$ on phase space
into operators ${\hat A}$ on Hilbert space, so enabling a Hilbert space formulation of 
classical mechanics \cite{Vercin00,Bracken03}.
We have
\bea
A_K(x,y)= {\cal W}^{-1}(A)_K(x,y)=    
\frac{1}{2\pi\hbar}\int    
A([x+y]/2,\,p)\,e^{ip(x-y)/\hbar}\,dp\,,    
\label{WW_inverse}
\eea
which defines the kernel of ${\hat A}$, and hence ${\hat A}$ itself, in terms of $A$.  In particular
the Groenewold operator ${\hat G}(t)$ is defined as 
${\hat G}={\cal W}^{-1}(2\pi\hbar\rho)$.
Then
\bea
\langle A\rangle (t)=\int A(q,\,p)\,dq\,dp={\rm Tr}({\hat A}{\hat G}(t))=\langle {\hat A}\rangle\,,
\mea\mea
{\rm Tr}({\hat G}(t))=1\,,\qquad\qquad\quad
\label{G_properties}
\eea
but ${\hat G}$ is not nonnegative definite in general; it is a quasidensity operator.

It can be seen that the development so far is completely analogous to the
development of the phase space formulation of quantum mechanics.  We can use ${\cal W}^{-1}$ to map all
of classical mechanics into a Hilbert space formulation \cite{Vercin00,Bracken03}. To complete the story,
we need to say what happens to the classical evolution of the Liouville density
\bea
\rho_t=H\,J\,\rho
\label{poisson}
\eea
under
the action of ${\cal W}^{-1}$.  Obviously the LHS maps into ${\hat G}_t$; the question is,
what happens to the Poisson bracket on the RHS.  

\section{The odot product and odot bracket}

To proceed, we note firstly by analogy with the definition of the star product that we can define a commutative odot
product of operators on Hilbert space

\bea
{\hat A}\odot{\hat B}={\cal W}^{-1}(AB)\,.
\label{odot_product}
\eea
This is an interesting product, quite distinct from the well known Jordan product of operators, which is also
commutative.   Unlike the Jordan product, however, this odot product is  associative.  Some of its
other characteristic properties have been described elsewhere \cite{Vercin00,Bracken03,Dubin04}.  

Next we note that
$A_q=\{A,\,p\}_{\star}$
so that
\bea
{\cal W}^{-1} (A_q)=\frac{1}{i\hbar}[{\hat p},\,{\hat A}]={\hat A}_q\,,\!\!\quad {\rm say},
\label{q_derivative1}
\eea
where ${\hat p}={\cal W}^{-1}(p)$.
Similarly, we define 
\bea
{\cal W}^{-1} (A_p)=\frac{1}{i\hbar}[{\hat A},\,{\hat q}]={\hat A}_p\,,
\mea\mea
{\cal W}^{-1} (A_{qp})
=\frac{1}{(i\hbar)^2}[[{\hat p},\,{\hat A}],\,{\hat q}]={\hat A}_{qp}\,,\quad etc.
\label{q_derivative2}
\eea

Then 
\bea
{\cal W}^{-1} \left( A\,J\,B\right)= {\hat A}_q \odot {\hat B}_p-{\hat A}_p\odot {\hat B}_q =
\frac{1}{i\hbar}[{\hat A},\,{\hat B}]_{\odot}\,,\quad \!\!{\rm say}\,,
\label{odot_bracket}
\eea 
which defines the odot bracket as the inverse image of $i\hbar\times$ the Poisson bracket.
From the Poisson bracket it inherits antisymmetry and a Jacobi identity.  

Our next task is to find an expansion of the odot bracket in 
ascending powers of $\hbar$, analogous to the expansion
(\ref{star_bracket1}) of the star bracket, in order to 
define a sequence of semiquantum approximations to classical dynamics.

\section{Semiquantum mechanics} 
We set
\bea
M=\frac{2}{\hbar}\sin\left( \frac{\hbar J}{2}\right)\,,
\label{moyal_operator}
\eea
so we can write for any two functions $A$ and $B$,
\bea
\{A,\,B\}_{\star}=A\,M\,B\,,
\label{moyal_operator2}
\eea
and note therefore that
\bea
{\cal W}^{-1}(A\,M\,B)=\frac{1}{i\hbar}[{\hat A},\,{\hat B}]\,.
\label{transform_moyal}
\eea
Next we write the Poisson bracket as 
\bea
A\,J\,B=A\,\frac{\hbar J/2}{\sin(\hbar J/2)}\,M\,B
\label{inverse_series1}
\eea
and, noting that 
\bea
\frac{\theta}{\sin(\theta)}=1+\theta^2/6+7\theta^4/360-\dots
\label{inverse_series2}
\eea
we obtain from (\ref{transform_moyal}) and (\ref{q_derivative2}) that
\bea
{\cal W}^{-1}(A\,J\,B)=\qquad\qquad\qquad\qquad\qquad\qquad\qquad
\mea\mea
\frac{1}{i\hbar}[{\hat A},\,{\hat B}]
-\frac{i\hbar}{24}\left([{\hat A}_{qq},\,{\hat B}_{pp}]
-2[{\hat A}_{qp},\,{\hat B}_{qp}]
+[{\hat A}_{pp},\,{\hat B}_{qq}]\right)+{\rm O}(\hbar^3)\,.
\label{poisson_expansion1}
\eea
Now we can answer the question raised at the end of Section 2.  Applying ${\cal W}^{-1}$ to
both sides of (\ref{poisson}), we obtain
\bea
{\hat G}_t=\qquad\qquad\qquad\qquad\qquad\qquad\qquad\qquad\qquad
\mea\mea
\frac{1}{i\hbar}[{\hat H},\,{\hat G}]
-\frac{i\hbar}{24}\left([{\hat H}_{qq},\,{\hat G}_{pp}]
-2[{\hat H}_{qp},\,{\hat G}_{qp}]
+[{\hat H}_{pp},\,{\hat G}_{qq}]\right)+{\rm O}(\hbar^3)\,.
\label{G_expansion1}
\eea
Thus we see the evolution of ${\hat G}$ in Hilbert space, 
which is equivalent to the classical evolution of 
the Liouville density $\rho$ in phase space, is given as a series in ascending powers of $\hbar$.
Keeping successively more terms in the series, we define a sequence of approximations to the classical evolution.
Note that the lowest order term is just the quantum evolution,
which now appears as the lowest order approximation to classical dynamics.  As we add more and more terms,
we have the possibility to explore the classical-quantum interface starting from the quantum side.  
This is completely complementary to 
what we normally do with semiclassical approximations to quantum mechanics. 

\section{Examples}

J.G. Wood and I have explored 
semiquantum and semiclassical approximations for simple nonlinear systems with one degree of freedom \cite{Wood05}.
Rather than 
Hamiltonians  of the form (\ref{hamiltonian1}), we considered
\begin{equation}    
H \,=\, E\sum_{k=0}^K b_k \left(H_0/E\right)^k\,, 
\label{hamiltonian_set}  
\end{equation}    
where $H_0$ is the simple harmonic oscillator Hamiltonian
\bea
H_0 \,=\, p^2/2m + m\omega^2q^2/2\,, 
\label{SHO_hamiltonian}
\eea  
and $E$, $b_k$ are constants.  These have the advantage that they 
are analytically tractable, but still show characteristic differences between
the classical and quantum evolutions \cite{Milburn86}. In particular we considered

\bea H_2=H_0^2/E\,,\qquad {\hat H}_2=
\mu \hbar\omega
\,({\hat N}^2 + {\hat N} + 1/4)\,,
\mea\mea
 H_3=H_0^3/E^2\,,\quad {\hat H}_3=
\mu ^2 \hbar\omega\,
({\hat N}^3 +3{\hat N}^2/2 +2{\hat N} + 3/4)\,,
\mea\mea
{\rm where}\quad{\hat H}_0=\hbar\omega\,({\hat N}+1/2)\,,\qquad\mu=\hbar\omega/E\,,\quad
\label{hamiltonians}
\eea
with ${\hat H}={\cal W}^{-1}(H)$ in each case, and ${\hat N}$ the usual 
oscillator number operator.  

\begin{figure}    
\centerline{\psfig{figure=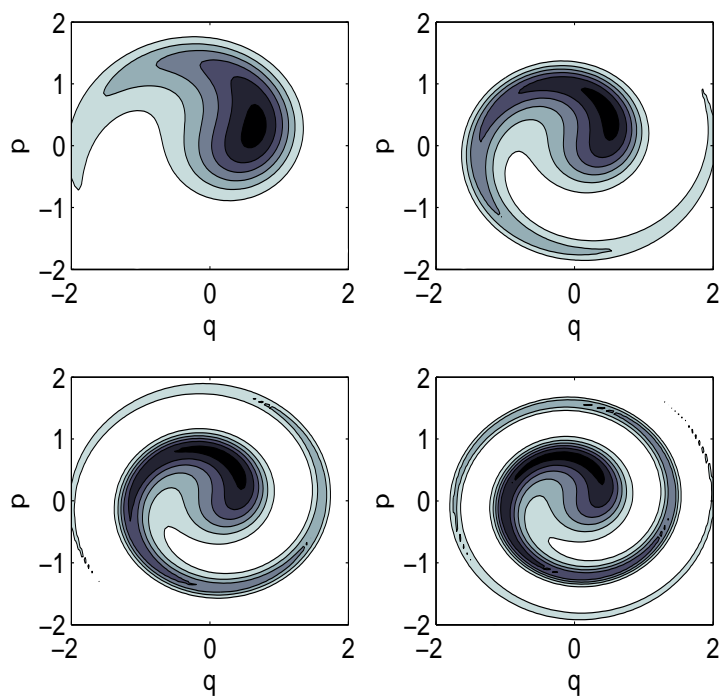,height=100mm,width=110mm}}    
\caption{Density plots showing the classical  
evolution of an initial Gaussian density centered at $q_0=0.5$, $p_0=0$ 
as generated by the Hamiltonian $H=H_0^2/E$.  
The parameters $m,\omega,E$ have been set equal to $1$, and the times of the plots are,
from left to right and top to bottom,  
$t=\pi/4$, $t=\pi/2$, $t=3\pi/4$ and $t=\pi$. 
\label{whorl_fig}  
} 
\end{figure}

\begin{figure}    
\centerline{\psfig{figure=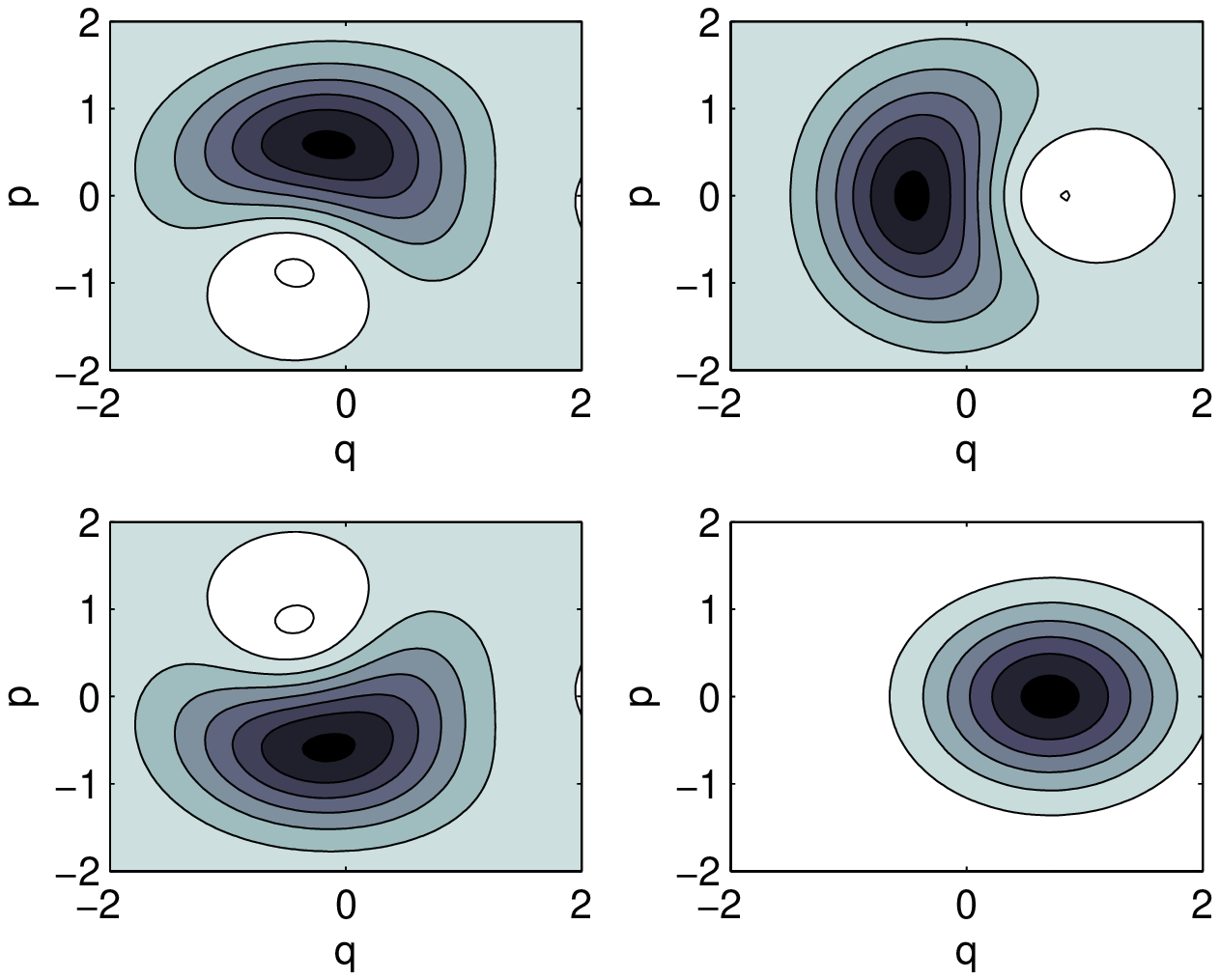,height=100mm,width=110mm}}    
\caption{Quantum evolution of  
an initial Gaussian Wigner function, with the same parameter values used  
in Figure \ref{whorl_fig}, and shown at the same times. Regions on which the
pseudo-density becomes negative are shown in white in the first three plots.
\label{quant_gaussian_fig}}  
\end{figure}

As an initial Liouville density on phase space, we took a Gaussian $\rho$ for which
the initial Groenewold operator ${\hat G}={\cal W}^{-1}(2\pi\hbar\rho)$ equals a true density operator, 
namely the density operator for a pure coherent state.  Differences between the classical and quantum evolutions of such
an initial state,
with the Hamiltonians
$H_2$ and ${\hat H}_2$, respectively,
are 
immediately apparent in the phase space plots Fig. \ref{whorl_fig} and Fig. \ref{quant_gaussian_fig}. 
Under the classical evolution, the density stays
positive everywhere, but develops ``whorls," whereas under the quantum evolution, 
the  density (Wigner function) becomes negative
on some regions (shown in white) and is periodic \cite{Milburn86}. 
Conversely, under the classical dynamics, the Groenewold operator ${\hat G}$  develops negative eigenvalues
\cite{Muga92,Muga93,Muga94,Habib02,Wood04},
whereas under the quantum evolution, such an initial pure-state 
density operator stays positive definite, with eigenvalues
$0$ and $1$. Simliar remarks apply with the Hamiltonians $H_3$ and ${\hat H}_3$.

There is no nontrivial semiquantum or semiclassical approximation ``in between"  
the classical and quantum dynamics for the Hamiltonians $H_2$ and ${\hat H}_2$.  Each of the series
(\ref{star_bracket1}) and (\ref{G_expansion1}) has just two terms in this case. 
For the series (\ref{star_bracket1}), the leading term 
defines the classical evolution, and adding the next term 
produces the full quantum dynamics.  Similarly, for the series (\ref{G_expansion1}), the 
leading term
defines the quantum evolution, and adding the next term produces the full
classical dynamics.  

\begin{figure}    
\centerline{\psfig{figure=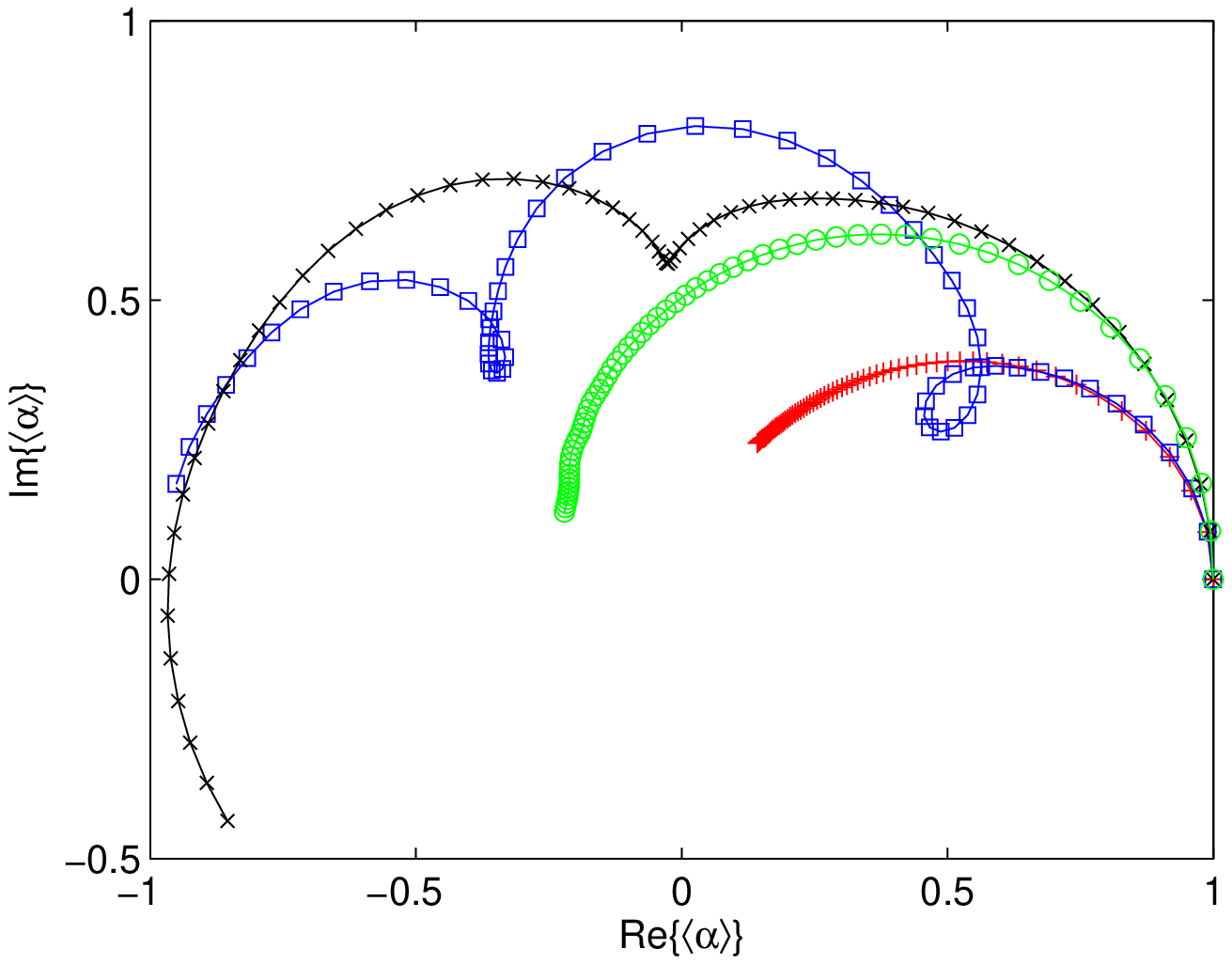,height=100mm,width=110mm}}    
\caption{Comparison of first moments of $\alpha=(q+ip)/\sqrt{2}$ for classical,  
semiquantum, quantum and semiclassical evolutions generated by the  
Hamiltonian $H=H_0^3/E^2$. Points on the classical, quantum, semiclassical  and semiquantum curves are  
labelled by +, x, o and $\square$, respectively.  
The evolution is  
over the time-interval $[0,\pi]$ and again $m=\omega=E=1$, with $\mu=1/2$, $\alpha_0=0.5$.
\label{cubic}}  
\end{figure}

To see interesting differences betweem semiclassical and semiquantum approximations, we considered
$H_3$ and ${\hat H}_3$, for which there are three terms in each of the series (\ref{star_bracket1}) and (\ref{G_expansion1}).  
Thus we can compare the classical evolution and  the first semiclassical approximation 
to quantum dynamics,
and also the quantum evolution and the first
semiquantum approximation to classical dynamics.
In Fig. \ref{cubic} we plot the expectation values of $q$, $p$ in the classical and semiclassical cases,
calculated at each time using the Liouville density $\rho$ or the Wigner function $W$ as in 
(\ref{averages1}), and also the expectation values of ${\hat q}$ and ${\hat p}$ in the quantum and semiquantum cases,
calculated at each time using the density operator ${\hat \rho}$ or the Groenewold operator ${\hat G}$ as in (\ref{G_properties}).  
From our
results it is already clear that semiquantum and semiclassical approximations provide different information 
about the interface between classical and quantum behaviours.

\begin{figure}    
\centerline{\psfig{figure=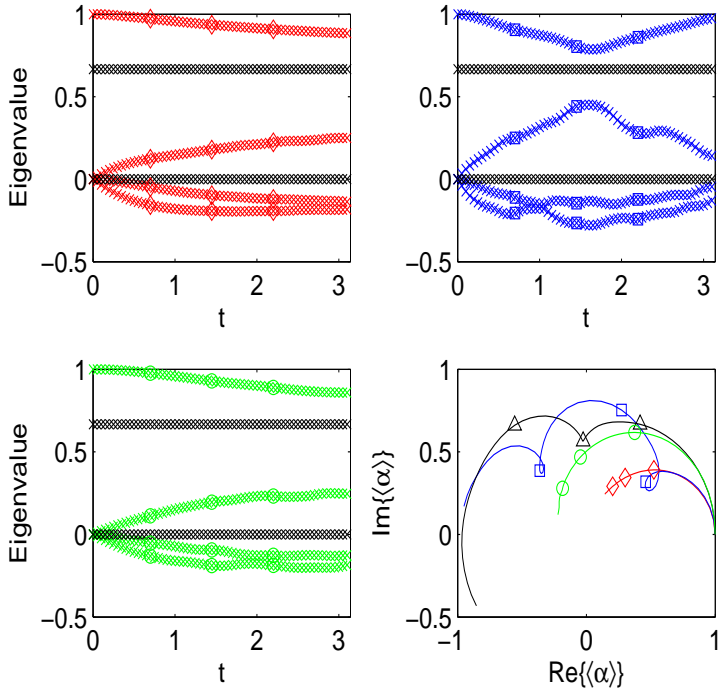,height=100mm,width=110mm}}    
\caption{Comparison of largest two and least two eigenvalues for, from
left to right and top to bottom, classical,  
semiquantum  and semiclassical evolutions generated by $H=H_0^3/E^2$,  
for the time interval $[0,\pi]$, and with $m=\omega=E=1$ and $\mu=1/2$, $\alpha_0=0.5$.  
The evolution of the first moment is reproduced from Fig. \ref{cubic} in the graph at bottom right for comparison.  
Each of the other graphs also features the quantum spectrum $\{0,1\}$ and in all graphs  
the values at the time-points $t=1,2,3$ are marked $\lozenge$ (classical), 
o (semiclassical), $\triangle$ (quantum) and $\square$ (semiquantum).  
\label{cubic_eigenvalues}}  
\end{figure}

These expectation values compare ``classical-like" properties of the different evolutions.  We also considered
``quantum-like" properties of the diferent evolutions, in particular the largest and smallest eigenvalues of
${\cal W}^{-1}(\rho)$ and ${\cal W}^{-1}(W)$ in the classical and semiclassical cases, and the largest and smallest eigenvalues of 
${\hat \rho}$ and ${\hat G}$ in the quantum and semiquantum cases.  The results are shown in Fig. \ref{cubic_eigenvalues}.

\section{Conclusions}

Semiquantum mechanics opens a new window on the interface between classical and quantum mechanics.  Our investigations
of examples as described above have not yet gone very far, but already we can say that semiquantum approximations show
characteristic differences from semiclassical approximations for a given nonlinear system.  More details are given in Ref. \cite{Wood05}. 

It is important to explore the nature of semiquantum approximations for other systems.  For example,
from the classical Hamiltonian $H$ as in (\ref{hamiltonian1}), we obtain 
\bea
{\hat H}=
{\hat p}^2/2m +V({\hat q})
\,, 
\label{hamiltonian2}
\eea
and substituting in (\ref{G_expansion1}), we get for the evolution of the Groenewold operator in such cases
\bea
{\hat G}_t= \frac{1}{i\hbar}[{\hat p}^2/2m +V({\hat q}),\,{\hat G}]
-\frac{i\hbar}{24}[V''({\hat q}),\,{\hat G}_{pp}]+{\rm O}(\hbar^3)\,,
\label{groenewold_evolution1}
\eea
which is to be compared with Wigner's famous formula (\ref{wigner_evolution1})
for the evolution of the Wigner function.  The implications of (\ref{groenewold_evolution1}) 
for various important choices of $V$, in particular
exactly solvable cases, should be examined.  

Of even more interest of course are systems with more degrees
of freedom that show chaos at the classical level.  
For such systems it should be particularly interesting to consider differences in the spectral
properties of ${\hat G}$ at different times in different approximations, in classically chaotic
or integrable regimes, as well as the different behaviours of expectation values of important phase space variables.   
 
We hope to investigate  some of these problems.

%


\section*{Acknowledgments}
Thanks to Dr. James Wood for much advice and assistance, and to 
a referee for bringing the paper by Vercin \cite{Vercin00} to our attention.
This work was supported by Australian Research Council Grant DP0450778.

\end{document}